\documentclass[11pt]{article}

\usepackage[a4paper,margin=2.6cm]{geometry}
\usepackage[T1]{fontenc}
\usepackage[utf8]{inputenc}
\usepackage{lmodern}
\usepackage{microtype}
\usepackage{amsmath,amssymb,mathtools}
\usepackage{braket}
\usepackage{bm}
\usepackage{hyperref}
\usepackage{enumitem}
\usepackage{booktabs}
\usepackage{amsthm}
\usepackage{tikz}
\usetikzlibrary{arrows.meta,positioning}
\usepackage{pgfplots}
\pgfplotsset{compat=1.16}

\hypersetup{
  colorlinks=true,
  linkcolor=blue,
  citecolor=blue,
  urlcolor=blue
}

\newcommand{\T}{\mathrm{T}}
\newcommand{\F}{\mathrm{F}}
\newcommand{\dec}{\mathrm{d}}
\newcommand{\s}{\mathrm{s}}
\newcommand{\HH}{\mathcal{H}}
\newcommand{\KK}{\mathcal{K}}
\newcommand{\Id}{\mathbb{I}}
\newcommand{\LL}{\mathcal{L}}
\newcommand{\Utr}{U_{\rm tr}}

\theoremstyle{plain}

\theoremstyle{remark}

\title{Revisiting the Aerts-Broekaert-Smets quantum model of the liar paradox}
\author{Massimiliano Sassoli de Bianchi\vspace{0.5 cm} \\ 
        Center Leo Apostel for Interdisciplinary Studies, 
         Brussels Free University \\ 
        \normalsize\itshape
         Krijgskundestraat 33, 1160 Brussels, Belgium \\
        \normalsize
        E-Mails: \url{msassoli@vub.ac.be},  \url{autoricerca@gmail.com}
          \vspace{0.5 cm}
              }
\date{September 07, 2026}

\begin{document}
\maketitle

\begin{abstract}
\noindent 
The quantum model of the two-sentence liar paradox proposed by Aerts, Broekaert, and Smets is an early example of the use of quantum formalism to describe cognitive dynamics. Our reconstruction is primarily pedagogical in intent, but it also leads to a number of clarifications, and to some new observations, concerning the structure of the model. Rewriting the model in Dirac notation, we make explicit the distinction between truth values originating from a decision and from semantic inference, and show that the associated enlargement of the one-sentence state space corresponds to a factorization $\mathcal H = \mathcal H_{\rm or}\otimes\mathcal H_{\rm tr}$, with respect to which the non-paradoxical configurations are separable while the genuine liar state is entangled.  We also derive the Hamiltonian generating the four-state cyclic evolution, express it in compact operator form, and give the resulting transition probabilities in closed form. We emphasize that the liar cycle admits a unitarily equivalent representation in the four-dimensional truth-only space, where the unmeasured liar state is separable, so the dimensional enlargement is not required by the unitary part of the dynamics; it is required, however, by the measurement structure. The enlargement is also required by the dynamics as soon as revision processes are admitted, in which states with identical truth content but different origins have different successors, the origin degree of freedom then acting as a minimal form of cognitive memory. We conclude by discussing the possible relationship between the model and more general deliberative processes.
\end{abstract}

\section{Introduction}

In 1999, Aerts, Broekaert and Smets proposed one of the earliest applications of the quantum formalism to a genuinely cognitive and semantic problem: the liar paradox \cite{Aerts1999a,Aerts1999b}. Their starting point was the observation that the truth behavior of a self-referential system of propositions is strongly contextual. Reading one of the sentences and making a hypothesis about its truth value changes the state of the whole semantic entity, in a way that can naturally be modeled as a quantum measurement. In the two-sentence liar paradox, this initial contextual interaction is followed by a deterministic sequence of semantic inferences, producing the familiar oscillation of truth assignments. Aerts et al. showed that this oscillatory behavior can be embedded in a unitary dynamics and described by a Schr\"odinger equation.
The construction was subsequently generalized by Broekaert, Aerts and D'Hooghe to configurations of $m$ mutually referring sentences \cite{Broekaert2006}, and has more recently been revisited from the standpoint of topos quantum logic \cite{Zhou2024}.

These works belong to the early developments that eventually contributed to what is now broadly referred to as \emph{quantum cognition} \cite{busemeyer2025}. Their interest is not limited to the formal analogy between binary truth values and two-level quantum systems. More importantly, they introduced a state-space description of a semantic entity, represented cognitive interaction as a state-changing measurement, and modeled the subsequent internal evolution of the entity through unitary dynamics. In this sense, the liar paradox provided an unusually simple setting in which superposition, contextuality, entanglement and dynamical evolution could all be brought together within a single cognitive model.\footnote{It is worth stressing that the object being modeled -- an oscillating sequence of provisional truth assignments generated by repeated application of a semantic rule -- is closely related to what, in the logical literature, is described by the \emph{revision theory of truth} \cite{Gupta1993}, where the liar is characterized precisely by the instability of its revision sequences, rather than by the assignment of a third truth value as in the fixed-point approach \cite{Kripke1975}. Connections between quantum-logical structures and revision-theoretic semantics have also been explored elsewhere \cite{Engesser2002}. The model discussed here can be read as a Hilbert-space implementation of the revision dynamics of the two-sentence liar, in which the revision steps are generated by a one-parameter unitary group.}

The original framework is conceptually rich, but some of its mathematical elements have been presented in a rather concise manner. In the non-paradoxical two-sentence cases, a two-dimensional truth space for each sentence is sufficient. In the genuine liar case, however, the local state space is enlarged from $\mathbb C^2$ to $\mathbb C^4$, with the explanation that truth and falsehood values originating from decision and from semantic inference must be distinguished \cite{Aerts1999a}. In the more detailed treatment of \cite{Aerts1999b}, the authors also motivate this dimensional increase by observing that the inferential dynamics becomes a four-step rather than a two-step process. A further and more compelling justification is also given: no state of the restricted space $\mathbb C^2\otimes\mathbb C^2$ is such that the four single-sentence truth and falsehood projectors map it onto four mutually orthogonal states, which is what is required if these four projections are to represent the four distinct phases of the inferential cycle.\footnote{This requirement was later turned into a general theorem by Broekaert, Aerts and D'Hooghe \cite{Broekaert2006}, who proved that an $m$-sentence liar configuration requires exactly $n=2m$ dimensions per sentence, hence a $(2m)^m$-dimensional Hilbert space, and who also observed that the resulting dynamics only spans a $2m$-dimensional subspace of it.} Nevertheless, the semantic meaning of the four basis states remains largely implicit, and the extensive use of explicit vector and matrix representations makes the underlying structure of the model less transparent than it can be.

The purpose of this article is primarily educational and is to present the Aerts-Broekaert-Smets construction in a way that makes its conceptual and mathematical structure more explicit and, consequently, more transparent. We first reformulate the model entirely in Dirac notation and interpret the four-dimensional one-sentence space as carrying two independent binary labels: the \emph{truth} value itself and the \emph{origin} of that value, namely decision (d) and semantic inference (s). This leads naturally to the factorization $\mathcal H= \mathcal H_{\rm or} \otimes \mathcal H_{\rm tr}$. Within this representation, the non-paradoxical cases are seen to be separable across this origin-truth partition, whereas the genuine liar state is entangled across it. More specifically, the liar state can be written as a coherent coupling between the two stable truth-correlation structures associated with the non-paradoxical cases.

We then isolate the four-dimensional invariant subspace generated by the inferential cycle and show that its discrete evolution is simply a cyclic unitary shift. This makes its diagonalization immediate in a discrete Fourier basis and allows the corresponding Hamiltonian to be derived in a compact operator form, without resorting to the cumbersome matrix manipulations used in the original articles. We formulate the evolution on the complete $16$-dimensional Hilbert space and identify a small inconsistency affecting the discrete evolution operator and the Hamiltonian reported in the historical papers.

Subsequently, we reconsider in which precise sense this enlarged representation is required. We show that, on the $4$-dimensional liar orbit, the origin labels can be consistently forgotten without losing any information about the successive truth assignments: the resulting truth-only description lives directly in $\mathbb C^2\otimes\mathbb C^2\simeq\mathbb C^4$ and is unitarily equivalent to the dynamics on the active sector of the Aerts et al.\ construction. The identification realizing this equivalence, however, intertwines the unitary evolution but \emph{not} the decision projectors, and this is exactly the content of the measurement-theoretic requirement recalled above. Thus, the $16$-dimensional space is not a dynamical necessity for the unitary part of the process, while it does remain a necessity for the representation of the cognitive interactions that initiate it.

We then ask under which circumstances the additional origin degree of freedom becomes necessary for the evolution as well. We show that this occurs when states having the same truth-value content, but different origins, can have different successors. A simple extension of the double-liar process, involving hypothesis revision, illustrates this possibility, producing a unitary dynamics that explores an $8$-dimensional subspace of the extended space.

Finally, we discuss a broader conceptual implication of the construction. The liar paradox provides an extreme but particularly transparent example of a cognitive process in which an initially actualized state modifies the semantic context in such a way that the state becomes unstable and generates a sequence of further transitions. This suggests a possible connection with \emph{deliberative} processes, in which provisional choices can trigger semantic consequences that either stabilize or destabilize them. In other words, the liar paradox model contains a minimal dynamic mechanism that could be relevant to a broader class of cognitive processes.

\section{One-sentence system}
\label{singleliar}

It is useful to begin with the simplest case, the single liar
\begin{equation}
S:\quad \text{Sentence $S$ is false}\nonumber
\end{equation}
whose truth value oscillates under repeated evaluation: if $S$ is true (T), then it is false (F), but if $S$ is false then it is telling the truth (T), and so on, so we have the sequence: 
\begin{equation}
\T \Rightarrow \F \Rightarrow \T \Rightarrow \F \Rightarrow \T \Rightarrow\cdots
\label{inf-cycle-single}
\end{equation}
Let $\HH_{\rm tr}=\operatorname{span}\{\ket{\T},\ket{\F}\}\simeq \mathbb C^2$ be the Hilbert space associated with sentence $S$, when only its truth value is retained. The inferential step is then implemented by the discrete unitary evolution operator $U_{\rm L}$, such that $U_{\rm L}\ket{\F}= \ket{\T}$ and $U_{\rm L}\ket{\T}= \ket{\F}$, given by
\begin{equation}
U_{\rm L}=\ket{\F}\!\bra{\T}+\ket{\T}\!\bra{\F}, \quad U_{\rm L}^2=\Id
\label{flip-flop}
\end{equation}
whose eigenvectors are $\ket{\chi_\pm}=\tfrac{1}{\sqrt2}(\ket{\T}\pm\ket{\F})$, with eigenvalues $\pm 1$. 

We can embed this discrete dynamics in a continuous unitary evolution by choosing a time interval $\tau$ between consecutive logical phases and seeking a Hamiltonian $H_{\rm L}$ such that
\begin{equation}
U_{\rm L}=U_{\rm L}(\tau)=e^{-iH_{\rm L}\tau}, \quad U_{\rm L}(t)=e^{-iH_{\rm L}t}
\end{equation}
We have the spectral representation
\begin{equation} 
H _{\rm L}=E_+\ket{\chi_+}\!\bra{\chi_+}+E_-\ket{\chi_-}\!\bra{\chi_-}
\label{energies-single} 
\end{equation}
and in view of $U_{\rm L}\ket{\chi_\pm}=\pm \ket{\chi_\pm}$, we can choose $E_+=0$ and $E_-=\frac{\pi}{\tau}$, hence
\begin{equation}
H_{\rm L}=\frac{\pi}{\tau} \ket{\chi_-}\!\bra{\chi_-} = \frac{\pi}{2\tau}\left(\Id-U_{\rm L}\right)
\end{equation}

In the basis $\{\ket{\T},\ket{\F}\}$, $U_{\rm L}$ and $H_{\rm L}$ can be represented by the $2\times 2$ matrices
\begin{equation}
U_{\rm L}=
\begin{bmatrix}
0 & 1\\
1 & 0
\end{bmatrix},\quad 
H_{\rm L}= \frac{\pi}{2\tau}
\begin{bmatrix}
\phantom{-}1 & -1\\
-1 & \phantom{-}1
\end{bmatrix}
\end{equation}

\begin{figure}[htbp]
\begin{center}
\includegraphics[width=12cm]{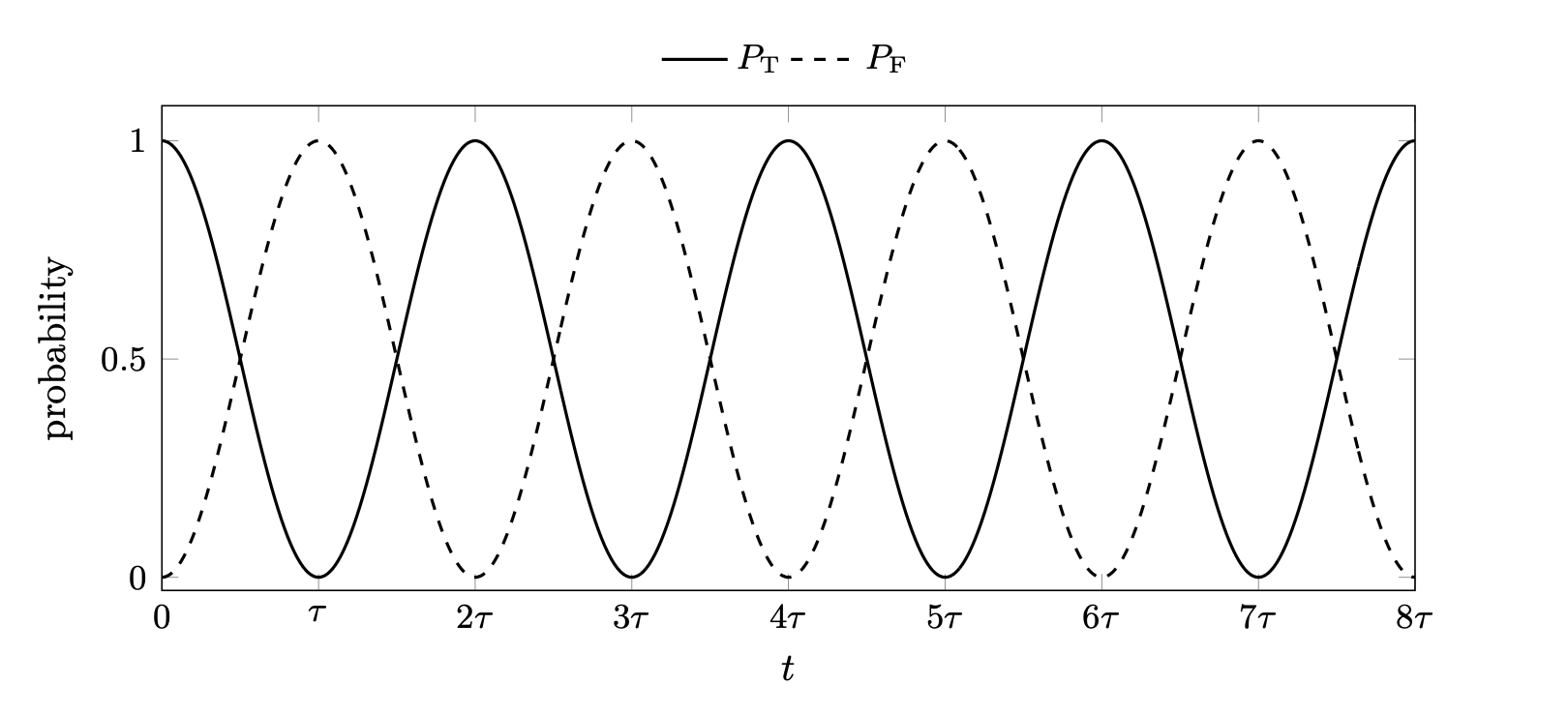}
\caption{The two transition probabilities for the single-sentence liar. At the discrete instants $t_n=n\tau$, $n=0,1,2,\dots$, the evolution, which is periodic with period $2\tau$, reproduces the inferential cycle (\ref{inf-cycle-single}).}
\label{figure1}
\end{center}
\end{figure}

So, the single liar lives in $\mathbb C^2$. If we consider $\ket{\chi_+}$ to be the unmeasured state, it is stationary, while each of the two eigenstates of the truth measurement is not, and evolves into the other after a time interval $\tau$. We can see this by considering the transition probabilities $P_{\T}(t)=|\braket{\T|U_{\rm L}(t)|\T}|^2$ and $P_{\F}(t)=|\braket{\F|U_{\rm L}(t)|\T}|^2= 1-P_{\T}(t)$. Writing $\ket{\T}=\tfrac{1}{\sqrt2}(\ket{\chi_+}+\ket{\chi_-})$, one gets $\braket{\T|U_{\rm L}(t)|\T}=\tfrac12(1+e^{-i\pi t/\tau})$, hence (see Fig.~\ref{figure1}): 
\begin{equation}
P_{\T}(t)=\cos^2\!\left(\tfrac{\pi t}{2\tau}\right),\quad P_{\F}(t)=\sin^2\!\left(\tfrac{\pi t}{2\tau}\right)
\label{probs-single}
\end{equation}

\section{Two-sentence systems}

In \cite{Aerts1999a,Aerts1999b} the following three pairs of mutually referring sentences are considered:
\begin{align}
(A)\quad
&S_1:\quad \text{Sentence $S_2$ is false}
\nonumber\\[-1mm]
&S_2:\quad \text{Sentence $S_1$ is true}\nonumber
\\[2mm]
(B)\quad
&S_1:\quad \text{Sentence $S_2$ is true}
\nonumber\\[-1mm]
&S_2:\quad \text{Sentence $S_1$ is true}\nonumber
\\[2mm]
(C)\quad
&S_1:\quad \text{Sentence $S_2$ is false}
\nonumber\\[-1mm]
&S_2:\quad \text{Sentence $S_1$ is false}\nonumber
\end{align}

Cases $(B)$ and $(C)$ admit globally consistent truth assignments. More precisely, case $(B)$ admits $(\T,\T)$ and $(\F,\F)$. Indeed, assuming $S_1$ is true, one obtains the sequence 
\begin{equation}
S_1=\T \Rightarrow S_2=\T \Rightarrow S_1=\T \Rightarrow S_2=\T \Rightarrow S_1=\T \Rightarrow\cdots
\end{equation}
Assuming instead that $S_1$ is false, one obtains the sequence 
\begin{equation}
S_1=\F \Rightarrow S_2=\F \Rightarrow S_1=\F \Rightarrow S_2=\F \Rightarrow S_1=\F \Rightarrow\cdots
\end{equation}

Case $(C)$ admits $(\T,\F)$ and $(\F,\T)$. Indeed, assuming that $S_1$ is true, one obtains the sequence
\begin{equation}
S_1=\T \Rightarrow S_2=\F \Rightarrow S_1=\T \Rightarrow S_2=\F \Rightarrow S_1=\T
\Rightarrow\cdots
\end{equation}
Assuming instead that $S_1$ is false, one obtains the sequence 
\begin{equation}
S_1=\F \Rightarrow S_2=\T \Rightarrow S_1=\F \Rightarrow S_2=\T \Rightarrow S_1=\F
\Rightarrow\cdots
\end{equation}

Case $(A)$, corresponding to the double liar paradox, is different, as truth assignments are not stable.  Indeed, assuming $S_1$ is true, one obtains the sequence
\begin{equation}
S_1=\T \Rightarrow S_2=\F \Rightarrow S_1=\F \Rightarrow S_2=\T \Rightarrow S_1=\T \Rightarrow\cdots
\end{equation}
In other words, we have the cycle
\begin{equation}
\rm{(T,F)\to (F,F)\to (F,T)\to (T,T)\to (T,F)}\to \cdots
\end{equation}
Similarly, assuming instead that $S_1$ is false, one obtains the sequence 
\begin{equation}
S_1=\F \Rightarrow S_2=\T \Rightarrow S_1=\T \Rightarrow S_2=\F \Rightarrow S_1=\F \Rightarrow\cdots
\end{equation}
corresponding to the equivalent cycle 
\begin{equation}
\rm{(F,T)\to (T,T)\to (T,F)\to (F,F)\to (F,T)}\to \cdots
\end{equation}
Thus, similarly to the single liar, the characteristic content of the double liar is that the truth assignments continue to oscillate and never stabilize.

\section{The non-paradoxical cases}

Let $\HH_{\rm tr}^{(i)} = \operatorname{span} \{\ket{\T},\ket{\F}\} \simeq \mathbb C^2$ be the minimal Hilbert space associated with sentence $i$, $i=1,2$, when only its truth value is retained. A generic normalized state is
\begin{equation}
\ket{\psi} = \alpha\ket{\T} + \beta\ket{\F}, \quad |\alpha|^2+|\beta|^2=1
\end{equation}
and truth and falsehood attributions are represented by the projectors
\begin{equation}
P_{\T} = \ket{\T}\!\bra{\T}, \quad P_{\F} = \ket{\F}\!\bra{\F},\quad P_{\T}+P_{\F}=\Id
\end{equation}
For two sentences, i.e., two quantum entities, the state space is then given by the tensor product  $\HH_{\rm tr}^{(1)}
\otimes \HH_{\rm tr}^{(2)} \simeq \mathbb C^2\otimes\mathbb C^2\simeq \mathbb C^{4}$. 

In case $(B)$, we observed that the only consistent truth assignments are $(\T,\T)$ and $(\F,\F)$. To describe this situation, Aerts, Broekaert and Smets used a maximally entangled state of the two sentences. We will adopt throughout the all-positive sign convention:\footnote{The two historical papers are not consistent here: in \cite{Aerts1999a} case $(B)$ is described by the antisymmetric combination $\tfrac{1}{\sqrt 2}(\ket{\T,\T}-\ket{\F,\F})$, whereas in \cite{Aerts1999b} the symmetric one is used; case $(C)$ is described by the singlet state in both. The relative sign is immaterial for the purpose at hand, as it amounts to a local phase transformation acting on one of the two sentences, $\ket{\T}\mapsto\ket{\T}$, $\ket{\F}\mapsto-\ket{\F}$, which leaves unaffected the correlation structure that these states are meant to encode.}
\begin{equation}
\ket{\Psi_B} = \tfrac{1}{\sqrt 2} \left(\ket{\T}\otimes\ket{\T} + \ket{\F}\otimes\ket{\F} \right)
\label{stateb}
\end{equation}

Within the pure-state representation adopted here, entanglement encodes the fact that $S_1$ and $S_2$ are not assigned independent
truth values but exhibit perfect correlations. For example, $P_{\T}\otimes \Id\ket{\Psi_B} \propto \ket{\T}\otimes\ket{\T}$, and $P_{\F}\otimes \Id\ket{\Psi_B} \propto \ket{\F}\otimes\ket{\F}$. Similarly, $\Id \otimes P_{\T}\ket{\Psi_B} \propto \ket{\T}\otimes\ket{\T}$ and $\Id \otimes P_{\F}\ket{\Psi_B} \propto \ket{\F}\otimes\ket{\F}$. 

In case $(C)$, we observed that the only consistent truth assignments are $(\T,\F)$ and $(\F,\T)$. To describe this situation, Aerts, Broekaert and Smets used a singlet state. Equivalently, we will use the corresponding Bell state with a positive sign:
\begin{equation}
\ket{\Psi_C} = \tfrac{1}{\sqrt 2} \left(\ket{\T}\otimes\ket{\F} + \ket{\F}\otimes\ket{\T} \right)
\label{statec}
\end{equation}
Again, within this pure-state representation, $S_1$ and $S_2$ are assigned perfectly anticorrelated truth values. For example, $P_{\T}\otimes \Id\ket{\Psi_C} \propto \ket{\T}\otimes\ket{\F}$, and $P_{\F}\otimes \Id\ket{\Psi_C} \propto \ket{\F}\otimes\ket{\T}$. Similarly, $\Id \otimes P_{\T}\ket{\Psi_C} \propto \ket{\F}\otimes\ket{\T}$, and $\Id \otimes P_{\F}\ket{\Psi_C} \propto \ket{\T}\otimes\ket{\F}$. 

So, these two cases can be easily represented in a truth-only two-sentence Hilbert space $\HH_{\rm tr}^{(1)}
\otimes \HH_{\rm tr}^{(2)}$, because the inferential propagation terminates in a consistent correlation.  In the liar paradox situation $(A)$, by contrast, the temporal order of ``assigned'' versus ``inferred'' truth becomes relevant if one wishes to retain explicitly the origin of each truth value. We shall see in Sec.~\ref{minimalspace}, however, that this additional bookkeeping is not required for a minimal representation of the four-state cycle itself, although it is required, as explained in Sec.~\ref{measurement}, for the representation of the decision projectors that initiate it.

\section{The liar paradox cycle}

A point that deserves to be made fully explicit is the semantic content of the dimensional doubling ($\mathbb C^2\longrightarrow\mathbb C^4$) given in \cite{Aerts1999b} for each sentence in case $(A)$. As recalled in the Introduction, the reason for this enlargement is that the four single-sentence projections of the initial state must yield four mutually orthogonal states. What the extra dimensions \emph{mean} becomes transparent if one distinguishes two questions: (i) What is the present truth value, $\T$ or $\F$? (ii) How did this truth value enter the present inferential step? Indeed, in the liar paradox cycle, a value can appear because the cognitive subject has just tested a given sentence, or because the value has been inferred from the other sentence. The same truth value, therefore, can  occur in two dynamically different roles. 

Accordingly, we introduce the four orthonormal one-sentence states $\ket{\T_{\dec}}$, $\ket{\F_{\dec}}$, $\ket{\T_{\s}}$ and $\ket{\F_{\s}}$, where the subscripts have the following meaning:  ``$\dec$'' is the truth value actualized following a decision, ``$\s$'' is the truth value obtained by semantic inference from the other sentence. Thus, the  one-sentence spaces are now 
\begin{equation}
\HH^{(i)} = \operatorname{span} \{\ket{\T_{\s}}, \ket{\F_{\s}}, \ket{\T_{\dec}}, \ket{\F_{\dec}}\} \simeq \mathbb C^4,\quad i=1,2
\label{Hilbert-i}
\end{equation}

This is the $m=2$ instance of the general labelling scheme later introduced in \cite{Broekaert2006}.\footnote{There, each sentence of an $m$-sentence configuration is given a $2m$-dimensional space whose basis states are read as ``sentence $i$ is true (respectively false) by inference after $j$ steps'', for $j\leq 2(m-1)$, together with ``sentence $i$ is true by hypothesis'' and ``sentence $i$ is false by hypothesis''. For $m=2$ there is a single inferential step, so the ``by inference'' states reduce to $\ket{\T_{\s}}$ and $\ket{\F_{\s}}$, while the ``by hypothesis'' states are our $\ket{\T_{\dec}}$ and $\ket{\F_{\dec}}$. The ordering adopted in (\ref{Hilbert-i}) is the one implicitly used in \cite{Aerts1999a,Aerts1999b}, where the truth and falsehood measurement projectors are placed on the third and fourth diagonal entries. We prefer the term \emph{decision} to ``hypothesis'' because, as noted below, only the first step of the process is a genuine hypothesis, all subsequent ones being forced endorsements of an inferred value.}
The two-sentence space is consequently $\HH = \HH^{(1)}\otimes\HH^{(2)} \simeq \mathbb C^4\otimes\mathbb C^4 \simeq \mathbb C^{16}$. This, however, should not be interpreted as the introduction of four truth values. There are only two truth values, $\T$ and $\F$. What has been doubled is the bookkeeping required to represent the \emph{origin} of a truth value in the dynamics. In that respect, a possible tensorial description of the state space associated with sentence $i$ is: $\HH^{(i)} \simeq \HH_{\rm or}^{(i)}\otimes\HH_{\rm tr}^{(i)}$, with $\HH_{\rm or}^{(i)} = \operatorname{span}\{\ket{\s},\ket{\dec}\}$, so that
\begin{equation}
\ket{\T_{\dec}} \equiv \ket{\dec}\otimes\ket{\T},\quad \ket{\F_{\dec}} \equiv \ket{\dec}\otimes\ket{\F},\quad \ket{\T_{\s}} \equiv \ket{\s} \otimes\ket{\T}, \quad \ket{\F_{\s}}\equiv \ket{\s} \otimes\ket{\F}
\end{equation}

\begin{figure}[ht]
\centering
\begin{tikzpicture}[>={Stealth[length=2.6mm]},node distance=3.1cm,
  every node/.style={font=\normalsize}]
\node (a0) at (0,2.4) {$\ket{a_0}=\ket{\T_{\dec},\F_{\s}}$};
\node (a1) at (5.6,2.4){$\ket{a_1}=\ket{\F_{\s},\F_{\dec}}$};
\node (a2) at (5.6,0)   {$\ket{a_2}=\ket{\F_{\dec},\T_{\s}}$};
\node (a3) at (0,0)   {$\ket{a_3}=\ket{\T_{\s},\T_{\dec}}$};
\draw[->] (a0) -- node[above,font=\normalsize]{$U$} (a1);
\draw[->] (a1) -- node[right,font=\normalsize]{$U$} (a2);
\draw[->] (a2) -- node[below,font=\normalsize]{$U$} (a3);
\draw[->] (a3) -- node[left,font=\normalsize]{$U$} (a0);
\end{tikzpicture}
\caption{The four-state inferential cycle of the double liar $(A)$. Reading the truth contents around the cycle, i.e., disregarding the origin labels, exactly one sentence changes its truth value at each step. The discrete evolution operator $U$ is defined in Section~\ref{evolutionsec}.}
\label{fig:cycle}
\end{figure}
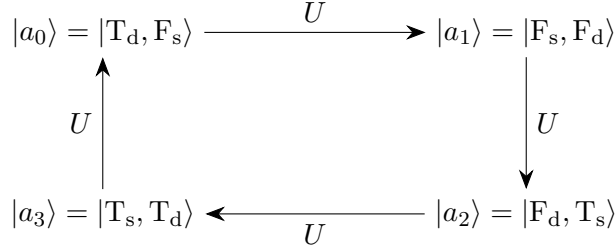

We can now use our state space representation to express the dynamics of the liar paradox. Suppose the cognitive interaction begins by selecting $S_1$ as true ($T_{\dec}$). Since $S_1$ says that $S_2$ is false, the semantic consequence is that $S_2$ is false ($F_{\s}$). This corresponds to the state
$\ket{a_0}=\ket{\T_{\dec}}\otimes\ket{\F_{\s}}\equiv\ket{\T_{\dec},\F_{\s}}$.
At this stage, a new cognitive interaction begins, in which the semantically inferred falsity of $S_2$ is taken as the selected truth value of $S_2$ ($F_{\dec}$). But $S_2$ asserts that $S_1$ is true. The falsity of $S_2$ therefore implies that $S_1$ is false ($F_{\s}$). This corresponds to the state
$\ket{a_1}=\ket{\F_{\s}}\otimes\ket{\F_{\dec}}\equiv\ket{\F_{\s},\F_{\dec}}$.
A third cognitive interaction then begins, in which the semantically inferred falsity of $S_1$ is taken as the selected truth value of $S_1$ ($F_{\dec}$). Since $S_1$ asserts that $S_2$ is false, $S_2$ becomes true ($T_{\s}$), which corresponds to the state
$\ket{a_2}=\ket{\F_{\dec}}\otimes\ket{\T_{\s}}\equiv\ket{\F_{\dec},\T_{\s}}$.
In the fourth cognitive interaction, the semantically inferred truth of $S_2$ is similarly taken as the selected truth value of $S_2$ ($T_{\dec}$); since $S_2$ asserts that $S_1$ is true, one infers that $S_1$ is also true ($T_{\s}$). This corresponds to the state
$\ket{a_3}=\ket{\T_{\s}}\otimes\ket{\T_{\dec}}\equiv\ket{\T_{\s},\T_{\dec}}$.
Finally, the semantically inferred truth of $S_1$ is taken as the selected truth value of $S_1$ ($T_{\dec}$), and the next inferential step returns to $\ket{a_0}$. Thus, we obtain the cycle represented in Fig.~\ref{fig:cycle}.

Coming to the initial, unmeasured state, Aerts, Broekaert and Smets choose the uniform superposition state
\begin{equation} 
\ket{\widetilde\Psi_A} = \tfrac{1}{2} \left(\ket{a_0} + \ket{a_1} + \ket{a_2} + \ket{a_3} \right)
= \tfrac{1}{2} \left( \ket{\T_{\dec},\F_{\s}} + \ket{\F_{\s},\F_{\dec}} + \ket{\F_{\dec},\T_{\s}} + \ket{\T_{\s},\T_{\dec}}\right)
\label{initialstate}
\end{equation}
which assigns equal amplitude to the four possible positions in the inferential orbit. Note that we have described the inference cycle using a decision process that starts with sentence $S_1$, but we could also have considered the first decision to be performed on $S_2$. Obviously, there is no reason to prefer one of these options over the other, nor to prefer one outcome of the decision process over another. It should also be noted that, although the d-label identifies decisions, distinguishing them from semantic inferences (s-label), such decisions can be either indeterministic (in the formulation of the initial hypothesis) or deterministic, when the semantic inference is transformed into an actualized choice.  

Before proceeding in the next section with the formal description of the evolution underlying the cycle represented in Fig.~\ref{fig:cycle}, we can observe that also the non-paradoxical cases $(B)$ and $(C)$ can be described using the extended Hilbert space (\ref{Hilbert-i}). The origin label then becomes redundant, and states (\ref{stateb})-(\ref{statec}) are replaced by:
\begin{align} 
\ket{\widetilde\Psi_B} &= \tfrac12 \left(\ket{\T_{\dec},\T_{\s}} + \ket{\T_{\s},\T_{\dec}} + \ket{\F_{\dec},\F_{\s}} + \ket{\F_{\s},\F_{\dec}} \right)\nonumber\\
\ket{\widetilde\Psi_C} &= \tfrac12 \left( \ket{\T_{\dec},\F_{\s}} + \ket{\T_{\s},\F_{\dec}} + \ket{\F_{\dec},\T_{\s}} + \ket{\F_{\s},\T_{\dec}} \right)
\label{extendedstates}
\end{align} 
Note that in these two cases the four states split into two disjoint two-step orbits, one associated with each of the consistent truth assignments, so that the space of states invariant under the corresponding discrete evolution is two-dimensional. The uniform superpositions (\ref{extendedstates}) are the symmetric choice within that space, expressing the absence of an a priori preference between the two consistent assignments.

The redundancy of the origin label in these two cases can be made explicit using the finer tensor-product representation introduced above, by reordering the factors in the Hilbert space tensorial decomposition as $\HH =(\HH_{\rm or}^{(1)}\otimes\HH_{\rm or}^{(2)})\otimes (\HH_{\rm tr}^{(1)}\otimes\HH_{\rm tr}^{(2)})$. Defining the normalized origin state
\begin{equation}
\ket{O}=\tfrac{1}{\sqrt2}\left(\ket{\s,\dec}+\ket{\dec,\s}\right)
\end{equation}
the two extended states (\ref{extendedstates}) factorize exactly as
\begin{equation}
\ket{\widetilde\Psi_B}=\ket{O}\otimes\ket{\Psi_B}, \quad
\ket{\widetilde\Psi_C}=\ket{O}\otimes\ket{\Psi_C}
\label{factorizationBC}
\end{equation}
Thus, in cases $(B)$ and $(C)$, the origin degree of freedom is separable from the truth-value degree of freedom, and tracing it out gives back the reduced states (\ref{stateb})-(\ref{statec}). This gives a precise sense in which the additional origin label is dynamically redundant in the non-paradoxical cases. By contrast, in case $(A)$ the origin and truth labels are entangled and we have instead: 
\begin{equation}
\ket{\widetilde\Psi_A} =\tfrac{1}{\sqrt2}\left(\ket{\s,\dec}\otimes \ket{\Psi_B} +\ket{\dec,\s}\otimes\ket{\Psi_C}\right)
\label{factorizationA}
\end{equation}
In other words, the liar paradox situation $(A)$ can be understood as a coherent coupling between the stable correlations of the $(B)$ and $(C)$ type, 
resulting from an entanglement across the origin-truth factorization.

\section{Discrete and continuous evolution}
\label{evolutionsec}

The discrete evolution operator $U$ describing the double liar paradox cycle is at this point simple to express, generalizing what we did in Section~\ref{singleliar}. It is useful first to isolate the dynamically relevant subspace
\begin{equation}
\KK=\operatorname{span}\{\ket{a_0},\ket{a_1},\ket{a_2},\ket{a_3}\}\subset\HH,\quad P_{\KK}=\sum_{n=0}^{3}\ket{a_n}\!\bra{a_n}
\label{Ksubspace}
\end{equation}
We know that $U\ket{a_n} = \ket{a_{n+1}}$, with $n$ defined modulo $4$. The action of the dynamics outside $\KK$ is not fixed by the inferential cycle, and we can choose the minimal extension in which it acts trivially on $\KK^\perp$. Thus, we can write
\begin{equation}
U=U_{\KK}+P_{\KK^\perp}
\label{UDextended}
\end{equation}
where $P_{\KK^\perp}= \Id-P_{\KK}$ is the projection operator onto $\KK^\perp$, and $U_{\KK}$ is given by
\begin{align} 
&U_{\KK} =  \ket{a_0}\!\bra{a_3}+\ket{a_1}\!\bra{a_0} + \ket{a_2}\!\bra{a_1} + \ket{a_3}\!\bra{a_2}\nonumber\\
&U_{\KK}^2=\ket{a_0}\!\bra{a_2}+\ket{a_1}\!\bra{a_3}+\ket{a_2}\!\bra{a_0}+\ket{a_3}\!\bra{a_1}= (U_{\KK}^2)^\dagger \nonumber\\
&U_{\KK}^3= \ket{a_0}\!\bra{a_1}+\ket{a_1}\!\bra{a_2}+\ket{a_2}\!\bra{a_3}+\ket{a_3}\!\bra{a_0}= U_{\KK}^\dagger \nonumber\\ 
&U_{\KK}^4=\ket{a_0}\!\bra{a_0}+\ket{a_1}\!\bra{a_1}+\ket{a_2}\!\bra{a_2}+\ket{a_3}\!\bra{a_3} =P_{\KK}
\label{ev}
\end{align}

Considering that $U_{\KK}^\dagger U_{\KK}=P_{\KK}$, we have $U^\dagger U = U_{\KK}^\dagger U_{\KK} + P_{\KK^\perp} = P_{\KK} + P_{\KK^\perp}=\Id$, and same for $UU^\dagger$, hence (\ref{UDextended}) is unitary. We can also observe that $U\ket{\widetilde\Psi_A}= \ket{\widetilde\Psi_A}$, i.e., that the initial state $\ket{\widetilde\Psi_A}$ is a stationary state of the discrete evolution. By contrast, each individual state $\ket{a_n}$ in the cycle is non-stationary, and this is why the liar's paradox arises, since the truth values never stabilize. 

In other words, the uniform superposition state is an invariant of the evolution, but when a decision selects one of the four states of the cycle, it subsequently evolves through the other states, in a progression that goes on forever. 

The discrete evolution $U$ can be easily diagonalized by the discrete Fourier basis. Let 
\begin{equation} 
\ket{\chi_k} = \frac12 \sum_{n=0}^{3} e^{i{\pi\over 2} nk}\ket{a_n}= \frac12 \left(\ket{a_0} +e^{i{\pi\over 2} k}\ket{a_1}  + e^{i{\pi\over 2} 2k}\ket{a_2}+e^{i{\pi\over 2} 3k}\ket{a_3}\right)
\end{equation}
with $k=0,1,2,3$. In view of (\ref{ev}), we have  
\begin{align} 
U \ket{\chi_k} &=  \tfrac12 \left(\ket{a_1} +e^{i{\pi\over 2} k}\ket{a_2}  + e^{i{\pi\over 2} 2k}\ket{a_3}+e^{i{\pi\over 2} 3k}\ket{a_0}\right)\nonumber\\
&= e^{-i{\pi\over 2} k}\tfrac12 \left(\ket{a_0}+e^{i{\pi\over 2} k}\ket{a_1} + e^{i{\pi\over 2} 2k}\ket{a_2}  + e^{i{\pi\over 2} 3k}\ket{a_3}\right)=e^{-i{\pi\over 2} k}\ket{\chi_k}
\label{evolution}
\end{align}
and of course $U \ket{\chi}=\ket{\chi}$ for all $\ket{\chi}\in \KK^\perp$. Note also that $\ket{\widetilde\Psi_A} = \ket{\chi_0}$, i.e., the invariance of the initial state is the statement that the zero Fourier mode has eigenvalue $1$.

As we did in Section~\ref{singleliar}, we can embed the discrete dynamics in a continuous unitary evolution by choosing a time interval $\tau$ between consecutive logical phases and seeking a Hamiltonian $H$ such that
\begin{equation}
U=U(\tau)=e^{-iH\tau}, \quad U(t)=e^{-iHt}, \quad U(n\tau)\ket{a_j}=\ket{a_{j+n}}
\label{discretecontinuous}
\end{equation}
where the indices are understood modulo $4$. In view of (\ref{evolution}), we have the spectral representation
\begin{equation} 
H =\sum_{k=0}^{3}E_k\ket{\chi_k}\!\bra{\chi_k},\quad E_k = \frac{\pi }{2}\frac{k}{\tau}
\label{energies}
\end{equation}
Note that no additional terms associated with $ \KK^\perp$ need to be considered, since the chosen branch assigns zero energy to that entire subspace.

Since $U^4=\Id$, every polynomial function of the one-step evolution operator $U=e^{-iH\tau}$ can be reduced to a polynomial of degree at most three. Moreover, since $U$ has the four distinct eigenvalues $\lambda_k=e^{-i{\pi\over 2} k}$, $k=0,1,2,3$, the operators $\Id,U,U^2,U^3$ are linearly independent and therefore form a basis of the four-dimensional commutative algebra generated by $U$. Since $H$ is obtained by spectral functional calculus as a function of $U$, we can therefore write
\begin{equation}
H=c_0\Id +c_1U+c_2U^2+c_3U^3
\end{equation}
This gives the four equations
\begin{equation}
c_0+c_1\lambda_k+c_2\lambda_k^2+c_3\lambda_k^3
=\tfrac{\pi k}{2\tau}, \quad k=0,1,2,3
\label{A1}
\end{equation}
Considering that $\lambda_0=1$, $\lambda_1=-i$, $\lambda_2=-1$ and $\lambda_3=i$, hence we need to solve
\begin{align}
&c_0+c_1+c_2+c_3 =0\qquad c_0-i c_1-c_2+i c_3 =\tfrac{\pi}{2\tau}\nonumber\\
&c_0-c_1+c_2-c_3 =\tfrac{\pi}{\tau}\qquad c_0+i c_1-c_2-i c_3 =\tfrac{3\pi}{2\tau}
\end{align}
Adding the four equations gives $4c_0=\tfrac{3\pi}{\tau}$, and therefore $c_0=\tfrac{3\pi}{4\tau}$. Adding the first and third equations yields $2(c_0+c_2)=\tfrac{\pi}{\tau}$, so that $c_2=-\tfrac{\pi}{4\tau}$. The remaining two coefficients can be obtained, for example, from $c_1+c_3=-\tfrac{\pi}{2\tau}$, and $c_3-c_1=i\tfrac{\pi}{2\tau}$, so in the end we get:
\begin{equation}
c_0=\tfrac{3\pi}{4\tau},\quad
c_1=-\tfrac{\pi}{4\tau}(1+i),\quad
c_2=-\tfrac{\pi}{4\tau},\quad
c_3=-\tfrac{\pi}{4\tau}(1-i)
\label{A2}
\end{equation}
Using $U^3=U^\dagger$, and observing that $(U^2)^\dagger= U^2$, the Hamiltonian takes the compact and manifestly self-adjoint form
\begin{equation}
H=\frac{\pi}{4\tau}
\left[3\, \Id-U^2-(1+i)U-(1-i)U^\dagger\right]
\label{hamiltonian2}
\end{equation}
Note that since $U$ acts as the identity operator on $\KK^\perp$, and  that $3-1-(1+i)-(1-i)=0$, $H$ acts as the null operator on $\KK^\perp$. On the other hand, on $\KK$, in the ordered basis $\{\ket{a_0},\ket{a_1},\ket{a_2},\ket{a_3}\}$, $U$ can be represented by the orthogonal $4\times 4$ matrix
\begin{equation}
U|_{\KK}=
\begin{bmatrix}
0 & 0 & 0 & 1\\
1 & 0 & 0 & 0\\
0 & 1 & 0 & 0\\
0& 0 & 1 & 0
\end{bmatrix}
\label{u-matrix}
\end{equation}
so that $H$ can be represented by the $4\times 4$ Hermitian matrix\footnote{The matrix (\ref{hamiltonianmatrix}) differs from the Hamiltonian reported in \cite{Aerts1999a,Aerts1999b}. Part of the difference is a legitimate choice of branch for the logarithm of a unitary operator, discussed below. The rest is of a different nature. First, in the ordered basis in which the four states of the cycle are there listed, the discrete evolution operator $U_D$ printed in \cite{Aerts1999a,Aerts1999b} implements the permutation $\ket{a_0}\to\ket{a_2}\to\ket{a_1}\to\ket{a_3}\to\ket{a_0}$, and not (\ref{u-matrix}); this is a discrepancy of basis ordering, implicitly acknowledged in \cite{Broekaert2006}, where the corresponding operator is given as a cyclic shift up to a basis permutation. Second, the Hamiltonian printed there neither commutes with $U_D$ nor admits $\ket{\widetilde\Psi_A}$ as an eigenvector, contrary to what is stated in the same papers. Changing the sign of its lower-right $2\times 2$ block restores both properties and yields exactly $\frac{2}{\pi}\, i\ln U_D$, with spectrum $\{-2,-1,0,1\}$, in agreement with the exponentials $e^{-it},e^{it},e^{2it}$ appearing in the evolution operator given there, and with the value $\tau=\pi/2$ used in their figures. Everything therefore points to a sign misprint.}
\begin{equation}
H|_{\KK}=\frac{\pi}{4\tau}
\begin{bmatrix}
3 & -1+i & -1 & -1-i\\
-1-i & 3 & -1+i & -1\\
-1 & -1-i & 3 & -1+i\\
-1+i & -1 & -1-i & 3
\end{bmatrix}
\label{hamiltonianmatrix}
\end{equation}

\begin{figure}[htbp]
\begin{center}
\includegraphics[width=12cm]{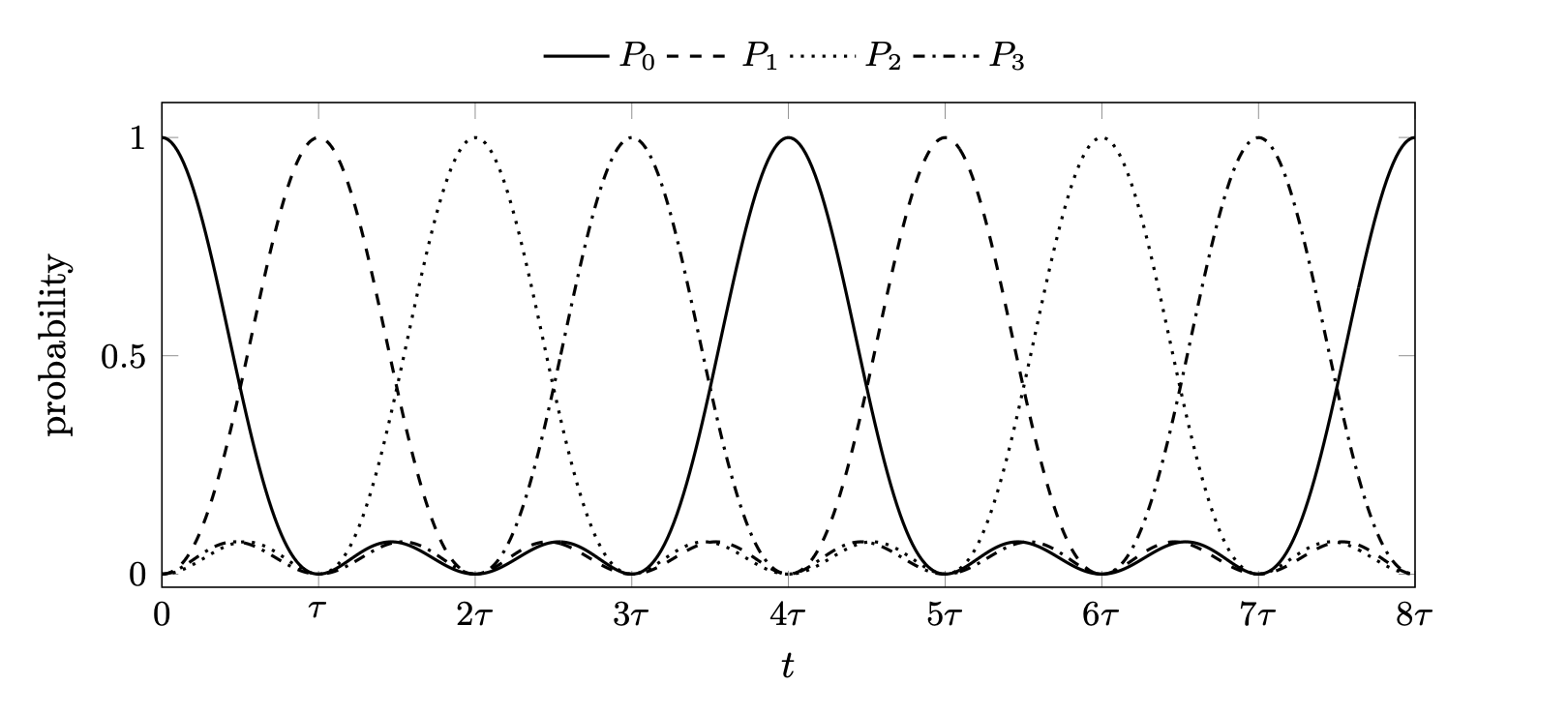}
\caption{The transition probabilities $P_n(t)=|\braket{a_n|U(t)|a_0}|^2$ for the double liar $(A)$, following a decision that has actualized the state $\ket{a_0}$ at $t=0$. At the discrete instants $t_n=n\tau$, $n=0,1,2,\dots$, the evolution, which is periodic with period $4\tau$, reproduces  the inferential cycle in Fig.~\ref{fig:cycle}.}
\label{figure2}
\end{center}
\end{figure}

So, we have obtained a continuous evolution $\ket{\Psi(t)}= U(t)\ket{\Psi(0)}$, in the Hilbert space $\HH \simeq\mathbb C^{16}$, where $\ket{\Psi(t)}$ obeys the time-dependent Schr\"odinger equation (with $\hbar=1$): 
\begin{equation} 
i\frac{d}{dt}\ket{\Psi(t)}=H\ket{\Psi(t)}
\end{equation}
If the initial condition is $\ket{\Psi(0)}= \ket{a_{j}}$ (depending on the outcome of the initial decision), it provides a continuous interpolation between the logical states, at the discrete instants $t_n=n\tau$ (see Figure~\ref{figure2}). By contrast, every initial state belonging to $\KK^\perp$ remains stationary.

The Hamiltonian is not uniquely determined, because the logarithm of a unitary operator is multivalued. In (\ref{energies}) we have adopted the simplest branch choice, also assigning zero energy to the entire sector outside the cyclic subspace. Different branches differ by integer multiples of $2\pi/\tau$ in the energies, and therefore generate exactly the same discrete dynamics at the instants $t_n=n\tau$, while the interpolation at intermediate times is in general branch-dependent. The choice (\ref{energies}) has the independent virtue of making $H$ positive semi-definite, with the unmeasured state $\ket{\widetilde\Psi_A}$ as a zero-energy ground state; within $\KK$ this ground state is non-degenerate, the residual degeneracy of the zero eigenvalue being entirely due to the inert sector $\KK^\perp$.

It is also worth observing that the spectrum (\ref{energies}) is equally spaced, and that the two bases $\{\ket{a_n}\}$ and $\{\ket{\chi_k}\}$ are mutually unbiased, since $|\braket{a_n|\chi_k}|^2=1/4$, for all $n,k$. The pair $(H,U)$ is therefore the Hamiltonian-shift pair of a four-level quantum clock, the states $\ket{a_n}$ playing the role of phase states and the states $\ket{\chi_k}$ that of energy states \cite{peres1980}. 

It is instructive to write down explicitly the transition probabilities $P_n(t)=|\braket{a_n|U(t)|a_0}|^2$, $n=0,1,2,3$. Expanding $\ket{a_0}$ and $\ket{a_n}$ in the Fourier basis, and using $\braket{a_n|\chi_k}=\tfrac12 e^{i\frac{\pi}{2}nk}$ together with (\ref{energies}), the amplitude reduces to a finite geometric sum:
\begin{equation}
\braket{a_n|U(t)|a_0}=\frac14\sum_{k=0}^{3}e^{-iE_kt}\,e^{i\frac{\pi}{2}nk}
=\frac14\sum_{k=0}^{3}e^{ik\theta_n(t)},\quad 
\theta_n(t)=\frac{\pi}{2}\left(n-\frac{t}{\tau}\right)
\label{amplitude}
\end{equation}
which can be summed in closed form. One obtains
\begin{equation}
P_n(t)=\frac{1}{16}\,\frac{\sin^2 2\theta_n(t)}{\sin^2\frac{\theta_n(t)}{2}}
=\frac{1}{16}\, \frac{\sin^{2}\!\left(\frac{\pi t}{\tau}\right)}
{\sin^{2}\!\left[\frac{\pi}{4}\left(n-\frac{t}{\tau}\right)\right]}
\label{probs-double}
\end{equation}
where at the isolated instants at which the denominator vanishes the expression is to be understood as its limit.\footnote{Equations (\ref{probs-single}) and (\ref{probs-double}) are the $N=2$ and $N=4$ instances of the same formula. For a cyclic shift of order $N$, with energies $E_k=\frac{2\pi}{N}\frac{k}{\tau}$, one finds $P_n(t)=\frac{1}{N^2}\sin^{2}\!\left[\frac{N}{2}\theta_n(t)\right]/\sin^{2}\!\frac{\theta_n(t)}{2}$, with $\theta_n(t)=\frac{2\pi}{N}\left(n-\frac{t}{\tau}\right)$, which for $N=2$ becomes $\cos^{2}\!\left[\frac{\pi}{2}\left(n-\frac{t}{\tau}\right)\right]$, i.e., (\ref{probs-single}).} Indeed, for $t=n\tau$ numerator and denominator vanish together and $P_n(n\tau)=1$, whereas $P_{n'}(n\tau)=0$ for $n'\neq n$: the four probabilities become equal to $1$ in turn, cyclically, thus reproducing the inferential cycle of Fig.~\ref{fig:cycle}, as they should. One also checks that $\sum_{n=0}^{3}P_n(t)=1$ at all times, the evolution remaining confined to $\KK$, and that the motion is periodic with period $4\tau$; see Figure~\ref{figure2}.

\section{The truth-only representation}
\label{minimalspace}

The preceding construction clarifies the meaning of the enlargement from $\mathbb C^2$ to $\mathbb C^4$ for each sentence: it preserves, in addition to the truth value itself, whether that value entered the current inferential step through decision or through semantic inference. Once this distinction has been made explicit, however, one can ask a more precise question: which part of the model actually requires the corresponding $16$-dimensional Hilbert space? We show in this section that the unitary evolution does not require it, whereas, as we will see in Sec.~\ref{measurement}, the measurement structure does.

The four states participating in the liar orbit, $\ket{\T_{\dec},\F_{\s}}$, $\ket{\F_{\s},\F_{\dec}}$, $\ket{\F_{\dec},\T_{\s}}$ and $\ket{\T_{\s},\T_{\dec}}$, 
carry four different truth assignments. If the origin labels are forgotten, they become, respectively, $\ket{\T,\F}$, $\ket{\F,\F}$, $\ket{\F,\T}$ and $\ket{\T,\T}$, which are precisely the four orthonormal basis states of the truth-only space $\HH_{\rm tr}^{(1)}\otimes\HH_{\rm tr}^{(2)}
\simeq \mathbb C^2\otimes\mathbb C^2\simeq\mathbb C^4$. Let us therefore define the \emph{truth-content map}
\begin{equation}
\Gamma:\HH\longrightarrow\HH_{\rm tr}^{(1)}\otimes\HH_{\rm tr}^{(2)},
\quad
\Gamma\ket{X_a,Y_{b}}=\ket{X,Y}, \quad
X,Y\in\{\T,\F\},\quad a,b\in\{\dec,\s\}
\label{forgetfulmap}
\end{equation}
extended by linearity. The map $\Gamma$ is linear and surjective, but it is not injective and not an isometry on $\HH$: for instance $\Gamma(\ket{\T_{\s},\F_{\s}}-\ket{\T_{\dec},\F_{\s}})=0$. Consider however its restriction to the dynamically relevant subspace $\KK$:
\begin{equation}
\Gamma|_{\KK}\ket{a_0}=\ket{\T,\F},\quad
\Gamma|_{\KK}\ket{a_1}=\ket{\F,\F},\quad
\Gamma|_{\KK}\ket{a_2}=\ket{\F,\T},\quad
\Gamma|_{\KK}\ket{a_3}=\ket{\T,\T}
\label{Gamma}
\end{equation}
which maps an orthonormal basis of $\KK$ onto an orthonormal basis of $\HH_{\rm tr}^{(1)}\otimes\HH_{\rm tr}^{(2)}$, and is therefore a unitary identification of the two spaces. The reduced cyclic evolution operator $\Utr$ and corresponding Hamiltonian $H_{\rm tr}$ then become
\begin{align}
&\Utr
=\ket{\T,\F}\!\bra{\T,\T} + \ket{\F,\F}\!\bra{\T,\F} +\ket{\F,\T}\!\bra{\F,\F} +\ket{\T,\T}\!\bra{\F,\T}
\nonumber\\
&H_{\rm tr}=\frac{\pi}{4\tau}
\left[3\Id-\Utr^2-(1+i)\Utr
-(1-i)\Utr^\dagger\right]
\label{Utruth}
\end{align}
and generate the cycle
\begin{equation}
\ket{\T,\F}\to\ket{\F,\F}\to\ket{\F,\T}
\to\ket{\T,\T}\to\ket{\T,\F}
\label{truthonlycycle}
\end{equation}
with $\Gamma|_{\KK}\,U|_{\KK}=\Utr\,\Gamma|_{\KK}$. In other words, the cyclic dynamics can be formulated directly in four dimensions, with no need to introduce additional degrees of freedom.

It is instructive to observe that the four ket-bra terms in (\ref{Utruth}) admit a more transparent rewriting. Inspection of the cycle (\ref{truthonlycycle}) shows that exactly one sentence changes its truth value at each step, and that the sentence which changes is determined by whether the two truth values presently agree or disagree. Introducing the projections
\begin{equation}
P_{=}=\ket{\T,\T}\!\bra{\T,\T}+\ket{\F,\F}\!\bra{\F,\F},\quad
P_{\neq}=\Id-P_{=}
\end{equation}
and with $T_{\rm flip}$ the truth-flip operator
\begin{equation}
T_{\rm flip}=\ket{\F}\!\bra{\T}+\ket{\T}\!\bra{\F}, \quad T_{\rm flip}^2=\Id
\end{equation}
which is the same as the evolution (\ref{flip-flop}), introduced in Sec.~\ref{singleliar} for the single sentence liar, one verifies directly that
\begin{equation}
\Utr=(T_{\rm flip}\otimes\Id)\,P_{\neq}+(\Id\otimes T_{\rm flip})\,P_{=},
\qquad
\Utr^{\,2}=T_{\rm flip}\otimes T_{\rm flip}
\label{condflip}
\end{equation}
That is: \emph{flip the first sentence if the two truth values differ, and the second sentence if they coincide}. This makes the mechanism of the paradox particularly explicit: the semantic injunctions of $(A)$ never allow the pair of truth values to settle, because whichever configuration is reached, the rule dictates a further flip.

Let us also consider how the reduction changes the way the initial state appears. Applying $\Gamma$ to Eq.~(\ref{initialstate}) gives
\begin{align}
\ket{\Psi_A}
&=\Gamma\ket{\widetilde\Psi_A}=\tfrac12\left(
\ket{\T,\F}+\ket{\F,\F}+\ket{\F,\T}+\ket{\T,\T}
\right)\nonumber\\
&=\tfrac{1}{\sqrt2}\left(\ket{\Psi_B}+\ket{\Psi_C}\right)
\label{reducedA}
\end{align}
where $\ket{\Psi_B}$ and $\ket{\Psi_C}$ are given by the entangled states (\ref{stateb}) and (\ref{statec}), respectively, whereas 
\begin{equation}
\ket{\Psi_A}
=\tfrac{1}{\sqrt2}\left(\ket{\T}+\ket{\F}\right)
\otimes
\tfrac{1}{\sqrt2}\left(\ket{\T}+\ket{\F}\right)
\label{separableA}
\end{equation}
is a product state.\footnote{Note that this is not an artifact of the all-positive sign convention adopted in (\ref{stateb})-(\ref{statec}). Up to a global phase, $\ket{\Psi_A}$ is the unique eigenvector of $\Utr$ with eigenvalue $1$, i.e., the unique state left invariant by the inferential dynamics, so that here the convention is selected by the dynamics rather than chosen by hand.} 
So, in the extended representation all three unmeasured states are maximally entangled across $S_1$ and $S_2$, whereas in the truth-only representation the two non-paradoxical states are maximally entangled and the liar state is separable.\footnote{Each of the three extended states (\ref{initialstate}) and (\ref{extendedstates}) is a sum of four terms occupying four distinct one-sentence states in each of the two factors; the associated $4\times4$ coefficient matrix therefore has exactly one non-vanishing entry, equal to $\tfrac12$, in each row and in each column, so that all four Schmidt coefficients across the $S_1|S_2$ bipartition are equal to $\tfrac12$. In the truth-only representation, by contrast, the Schmidt coefficients are $(\tfrac{1}{\sqrt2},\tfrac{1}{\sqrt2})$ for $\ket{\Psi_B}$ and $\ket{\Psi_C}$, and $(1,0)$ for $\ket{\Psi_A}$.} There is no contradiction, because $\Gamma|_{\KK}$, although unitary, is not a local transformation: it is not of the form $V_1\otimes V_2$, and it therefore has no reason to preserve the sentence-sentence bipartition.

\section{The measurement structure}
\label{measurement}

We now come to the point that explains the necessity of the dimensional enlargement. Consider the decision projectors. In the extended representation, a decision that assigns a truth value to a sentence produces, by definition, a value carrying the label $\dec$, so that the relevant projectors are $P_{\T_{\dec}}\otimes\Id$, $P_{\F_{\dec}}\otimes\Id$, $\Id\otimes P_{\T_{\dec}}$ and $\Id\otimes P_{\F_{\dec}}$, with $P_{X_{\dec}}=\ket{X_{\dec}}\!\bra{X_{\dec}}$. Applying them to the unmeasured state (\ref{initialstate}) one obtains $P_{\T_{\dec}}\otimes\Id\ket{\widetilde\Psi_A}\propto\ket{a_0}$, $\Id\otimes P_{\F_{\dec}}\ket{\widetilde\Psi_A}\propto\ket{a_1}$, $P_{\F_{\dec}}\otimes\Id\ket{\widetilde\Psi_A}\propto\ket{a_2}$, and $\Id\otimes P_{\T_{\dec}}\ket{\widetilde\Psi_A}\propto\ket{a_3}$, that is, exactly the four mutually orthogonal states of the cycle, each with probability $1/4$. 

This is precisely the property that the enlargement was introduced to secure. Indeed, there is no state $\ket{\Psi}\in \HH_{\rm tr}^{(1)}\otimes\HH_{\rm tr}^{(2)}$ such that the four states $P_{\T}\otimes\Id\ket{\Psi}$, $P_{\F}\otimes\Id\ket{\Psi}$, $\Id\otimes P_{\T}\ket{\Psi}$ and $\Id\otimes P_{\F}\ket{\Psi}$ are non-zero and mutually orthogonal. To show this, we write $\ket{\Psi}=\alpha\ket{\T,\T}+\beta\ket{\T,\F}+\gamma\ket{\F,\T}+\delta\ket{\F,\F}$. Then $P_{\T}\otimes\Id\ket{\Psi}=\alpha\ket{\T,\T}+\beta\ket{\T,\F}$ and $\Id\otimes P_{\T}\ket{\Psi}=\alpha\ket{\T,\T}+\gamma\ket{\F,\T}$, so that their scalar product equals $|\alpha|^2$; similarly, the scalar product of $P_{\T}\otimes\Id\ket{\Psi}$ with $\Id\otimes P_{\F}\ket{\Psi}=\beta\ket{\T,\F}+\delta\ket{\F,\F}$ equals $|\beta|^2$. Mutual orthogonality would then force $\alpha=\beta=0$, hence $P_{\T}\otimes\Id\ket{\Psi}=0$.

This is the $m=2$ case of the general result of \cite{Broekaert2006}, and it is the argument invoked in \cite{Aerts1999b} to justify the passage to $\mathbb C^4\otimes\mathbb C^4$. Concretely, applying a single-sentence truth projector to the reduced state (\ref{separableA}) gives
\begin{equation}
P_{\T}\otimes\Id\ket{\Psi_A}\propto\ket{\T}\otimes\tfrac{1}{\sqrt2}\left(\ket{\T}+\ket{\F}\right)
=\tfrac{1}{\sqrt2}\left(\Gamma\ket{a_0}+\Gamma\ket{a_3}\right)
\label{ambiguity}
\end{equation}
which is not one of the four states of the cycle, but a coherent superposition of the two phases of the cycle in which $S_1$ is true. Thus, the $16$-dimensional space of Aerts, Broekaert and Smets is a \emph{semantically enriched representation}, which is not required by the unitary part of the dynamics -- this part being carried by a four-dimensional truth-only space -- but which is required as soon as one wants the cognitive interactions themselves to be represented as projections producing the phases of the cycle.

Eq.~(\ref{ambiguity}) also has a direct semantic reading. It does not mean that, after a decision on $S_1$, the truth value of $S_2$ remains undetermined: the semantic injunction of $(A)$ fixes it immediately, since if $S_1$ is true then $S_2$ is false. What remains undetermined is \emph{which phase of the cycle} one is in, because the truth value of a single sentence does not identify it: both $\ket{a_0}$, where $S_1$ has just been decided to be true, and $\ket{a_3}$, where the truth of $S_1$ has been inferred from a decision on $S_2$, are compatible with $S_1$ being true. It is exactly this ambiguity that the origin label removes, and (\ref{ambiguity}) expresses the fact that the truth-only description cannot remove it. In cases $(B)$ and $(C)$, by contrast, no such ambiguity arises, because there the two phases compatible with a given truth value of one sentence carry the same truth assignment for the other sentence as well.

\section{When the origin becomes relevant}
\label{origin-relevance}

We emphasized that the dimensional enlargement introduced by Aerts, Broekaert and Smets is not necessary for describing the unitary evolution of the standard double-liar dynamics. This  raises a natural question: under which circumstances would the distinction between a truth value obtained through decision and one obtained through semantic inference become necessary for the evolution as well? 

Suppose that two extended states $\ket{\phi}$ and $\ket{\chi}$ have the same truth-value content, $\Gamma\ket{\phi}=\Gamma\ket{\chi}$, but the evolution assigns them different truth-value successors, $\Gamma U'\ket{\phi}\neq\Gamma U'\ket{\chi}$, where $U'$ denotes the relevant evolution operator, distinct from the four-cycle operator $U$ of Sec.~\ref{evolutionsec}.  In this case, no deterministic evolution operator acting on the truth-only space can reproduce the full dynamics. Indeed, the same truth-only state would have to possess two different successors. The origin degree of freedom then ceases to be redundant: it carries information about the previous history of the cognitive process that is required to predict its subsequent evolution. In this sense, the origin label can be interpreted as a minimal form of cognitive memory and the enlarged origin-truth space provides a representation in which part of the dynamically relevant history can be encoded in each state. 

To illustrate how this can occur, we consider a simple generalization of the double-liar process. In the cycle considered in the previous sections, each semantically inferred value was taken as the selected value initiating the next cognitive interaction. Thus, for example, the inference $F_{\s}$ for $S_2$ was subsequently taken as $F_{\dec}$ when $S_2$ became the sentence under consideration. This repeated transition from an inferred to a selected value is what produces the four-state cycle represented in Fig.~\ref{fig:cycle}. 

Consider instead a cognitive subject who seeks, at least initially, to avoid repeating the same four-step  cycle. The subject starts from the unmeasured state
$\ket{\widetilde\Psi_A}$ and makes an initial decision (hypothesis), say that $S_1$ is true, $\ket{\widetilde\Psi_A} \rightarrow \ket{\T_{\dec},\F_{\s}}$, then follows the pure semantic consequences of this hypothesis:
\begin{equation}
\ket{\widetilde\Psi_A} \rightarrow \ket{\T_{\dec},\F_{\s}} \rightarrow \ket{\F_{\s},\F_{\s}} \rightarrow \ket{\F_{\s},\T_{\s}} \rightarrow \ket{\T_{\s},\T_{\s}} \rightarrow \ket{\T_{\s},\F_{\s}}\rightarrow\cdots
\label{seq1}
\end{equation}
Note that, apart from the initial decision, all labels in this first inferential sequence are now s-labels. States such as $\ket{\F_{\s},\F_{\s}}$, in which both truth values carry the inference label, do not occur in the standard cycle, where each state carries exactly one d-label and one s-label. They are nevertheless meaningful in the present protocol: they describe a configuration in which both truth values have been reached by inference from the initial hypothesis, without any of them having been endorsed as a new decision. 

Once it becomes apparent that the reasoning is returning to a configuration already encountered, this time the cognitive subject does not simply reselect the same inferred truth value and repeat the cycle. Instead, the recurrence is taken as evidence that the current hypothesis has failed to produce a stable truth assignment and that a different hypothesis needs to be actively selected. More precisely, continuing the semantic inference in (\ref{seq1}) would produce the transition  $\ket{\T_{\s},\F_{\s}}\rightarrow \ket{\F_{\s},\F_{\s}}$, which is an already encountered state. 

An alternative value for $S_1$ is then selected, producing the state $\ket{\F_{\dec},\T_{\s}}$. The semantic consequence of this new selection is the state $\ket{\T_{\s},\T_{\s}}$, which is again a configuration that has already been encountered, hence the cognitive subject will 
again avoid repeating it. This time the subject will tentatively change the truth value of the second sentence. This gives the state  $\ket{\F_{\s},\F_{\dec}}$, the semantic consequence of which is the state $\ket{\F_{\s},\T_{\s}}$. The latter being an already encountered state, the next alternative is $\ket{\T_{\s},\T_{\dec}}$, which would also lead to the already encountered state $\ket{\T_{\s},\F_{\s}}$. At this point, the subject would have to change again the truth value of the first sentence, which brings the process back to $\ket{\F_{\dec},\T_{\s}}$, i.e., to a state already visited. In other words, the continuation of (\ref{seq1}) is: 
\begin{equation}
\cdots \rightarrow \ket{\F_{\dec},\T_{\s}} \rightarrow \ket{\F_{\s},\F_{\dec}} \rightarrow \ket{\T_{\s},\T_{\dec}} \rightarrow\cdots 
\label{seq2}
\end{equation}

The process just described involves eight mutually orthogonal basis states, which it is convenient to relabel as
\begin{align}
&\ket{b_0}=\ket{\T_{\dec},\F_{\s}},\quad
\ket{b_1}=\ket{\F_{\s},\F_{\s}},\quad
\ket{b_2}=\ket{\F_{\s},\T_{\s}},\quad
\ket{b_3}=\ket{\T_{\s},\T_{\s}},\nonumber\\
&\ket{b_4}=\ket{\T_{\s},\F_{\s}},\quad
\ket{b_5}=\ket{\F_{\dec},\T_{\s}},\quad
\ket{b_6}=\ket{\F_{\s},\F_{\dec}},\quad
\ket{b_7}=\ket{\T_{\s},\T_{\dec}}
\label{eightstates}
\end{align}
and it therefore unfolds in an $8$-dimensional subspace $\LL\subset\HH$, which contains $\ket{\widetilde\Psi_A}$, since $\ket{a_0}=\ket{b_0}$, $\ket{a_1}=\ket{b_6}$, $\ket{a_2}=\ket{b_5}$ and $\ket{a_3}=\ket{b_7}$. The last step of the protocol, from $\ket{b_7}$, is the one where the subject would have to revise the truth value of the first sentence. If this revision is performed by endorsing the value of $S_1$ inferred in $\ket{b_7}$, one returns to $\ket{b_0}$ and the whole process closes on itself, so that the revision dynamics is the cyclic shift $U'\ket{b_n}=\ket{b_{n+1}}$,  $U'^{\,8}=\Id|_{\LL}$, with $n$ defined modulo $8$, which is unitary on $\LL$ and can be diagonalized exactly as in Sec.~\ref{evolutionsec}, replacing the fourth roots of unity by the eighth roots of unity, and its unique invariant state is the uniform superposition $\tfrac{1}{\sqrt8}\sum_{n=0}^{7}\ket{b_n}$. Note that this is different from $\ket{\widetilde\Psi_A}$: the enlarged process has its own indeterminate state.

An alternative reading, closer to the description given in \cite{Aerts1999b,Broekaert2006}, is that when reaching state $\ket{b_7}=\ket{\T_{\s},\T_{\dec}}$, the cognitive subject recognizes that the attempted revisions are themselves returning to previously explored states, abandons the attempt to establish a stable truth assignment, and terminates the cognitive interaction, the conceptual entity thereby returning to the initial superposition $\ket{\widetilde\Psi_A}$. Note however that this last step $\ket{b_7}\rightarrow\ket{\widetilde\Psi_A}$ would not be unitary. In this reading, the process must therefore be described in three stages: an indeterministic initial decision, a unitary evolution on $\LL$, and a final non-unitary release of the entity back to its unmeasured state, the latter representing the termination of the cognitive context rather than a further inferential step.

In either reading, the point of interest is the same. The truth contents of the eight states (\ref{eightstates}) are, in order,
\begin{equation}
(\T,\F)\to(\F,\F)\to (\F,\T)\to (\T,\T)\to (\T,\F)\to (\F,\T)\to (\F,\F)\to (\T,\T)
\end{equation}
so that $\Gamma$ maps $\LL$ two-to-one onto the truth-only space, each truth configuration occurring exactly twice. Two occurrences can now have different successors. For instance $\Gamma\ket{b_0}=\Gamma\ket{b_4}=\ket{\T,\F}$, but $\Gamma U'\ket{b_0}=\ket{\F,\F}\neq \ket{\F,\T}=\Gamma U'\ket{b_4}$. Hence, unlike in the standard double-liar cycle, truth assignments alone no longer determine the evolution, and a truth-only representation in $\mathbb C^4$ is no longer sufficient. Two states that become identical under $\Gamma$ now have different cognitive meanings and different successors.

A remark is in order. The rule that generates (\ref{seq1})--(\ref{seq2}) is, as stated, a history-dependent rule, and a history-dependent rule does not in general define a map on states at all, but only on pairs consisting of a state and a record of the states already visited. What makes the present example well posed is that, along the trajectory considered, each of the eight states occurs only once, so that the rule collapses into an ordinary, memoryless map on $\LL$. This is a property of the particular protocol, but more elaborate revision strategies would require the record itself to be part of the state space, i.e., a further enlargement. The origin label can then be seen as the first and simplest step in that direction.

\section{Concluding remarks}

In our revisitation of the two-sentence liar paradox proposed by Aerts, Broekaert, and Smets, we have clarified that their extended $16$-dimensional construction is a \emph{semantically enriched representation}: it records not only which truth values are present but also how they entered the inferential process. The $4$-dimensional truth-only construction is instead a \emph{minimal dynamical representation}: it retains exactly the information needed to identify the four phases of the liar cycle and their unitary evolution.

Whether in the extended or minimal representation, the initial state of the bipartite conceptual entity is represented by a uniform superposition of
the four states relevant to the description, reflecting the absence of any a priori preference for a particular initial truth-value
actualization. Hence, the initial state describes a situation of equiprobability. It is also a stationary state, but once its symmetry is broken by the actualization of a first truth value, for one of the two sentences, the pre-decision state changes according to the quantum projection postulate, becoming one of the four possible states in the cycle represented in Fig.~\ref{fig:cycle}. In the extended representation, this produces one of the transitions
\begin{equation}
\ket{\widetilde\Psi_A}\longrightarrow\ket{a_0}, \ket{a_1}, \ket{a_2},\text{ or } \ket{a_3}
\end{equation}
each with probability $1/4$. In the minimal representation, the corresponding transitions
\begin{equation}
\ket{\Psi_A}\longrightarrow
\ket{\T,\F},\ket{\F,\F},\ket{\F,\T},\text{ or }\ket{\T,\T}
\end{equation}
are also obtained with probability $1/4$ each, but correspond to a measurement associated with the complete orthonormal basis of the truth-only space, that is, to a joint reading of the truth values of both sentences, and not to a projector acting on a single sentence. As we have seen in Eq.~(\ref{ambiguity}), a single-sentence projector would leave the entity in a superposition of two phases of the cycle. This is the precise price of the reduction: the truth-only representation describes the same unitary dynamics with the same probabilities, but it can no longer represent the cognitive act that initiates it as an interaction with a single sentence.

We conclude our analysis of the historical model by Aerts et al. with a more general observation about decision-making processes, which can also include relatively deterministic reflective phases in which a cognitive entity explores the consequences of a provisional choice before reaching a final resolution. The quantum dynamics underlying the liar paradox provides an extreme example of a cognitive process in which, once a particular alternative is provisionally selected, for example as a working hypothesis, this selection activates a trajectory of meanings that may eventually be interrupted by a final choice. This suggests an interesting correspondence with the more general phenomenon of deliberation.

We can distinguish \emph{deliberation} from \emph{choice}, representing the latter as the selection of one among different alternatives, whereas the former refers to the consideration of reasons, consequences, and trade-offs associated with those alternatives. In the models considered above, the distinction between a provisional choice that consolidates and one that undermines itself admits a simple formulation. A choice actualizes a state $\ket{\psi}$ and, at the same time, activates a context that determines a subsequent evolution $V$. The choice is \emph{self-reinforcing} when $\ket{\psi}$ is an eigenstate of $V$, so that the deliberation returns the entity to the same truth assignment, and \emph{self-destabilizing} when it is not, so that the deliberation drives the entity away from it. Cases $(B)$ and $(C)$ realize the first situation, case $(A)$ the second.

What makes the liar paradox extreme is that the context is a purely logical injunction, so that the transition is deterministic and the destabilization never terminates; in more realistic situations, the activated context is itself indeterministic. This is because the selection of a response reorganizes the entire semantics of the situation, and the associations thus activated can either provide further support for that response, leading to its stabilization, or increase the plausibility of an alternative response, leading to a sequence of further choices.

Consider the following simple example. A speaker makes one unusually long pause during an otherwise fluent speech, and suppose that a listener is asked whether the pause was accidental or deliberate. Imagine that the listener provisionally selects `accidental' as a working hypothesis. From this same initial choice, two quite different deliberative processes may develop, depending on which aspects of the context become salient.

In a self-destabilizing process, considering the pause as accidental may draw attention to the fact that it occurred at exactly the rhetorically appropriate moment, just before an important statement. This contextual feature is difficult to reconcile with a purely accidental hesitation and therefore increases the plausibility of the alternative answer, `deliberate'. If the listener now adopts `deliberate' as the new working hypothesis, other features may in turn become salient, for instance, that the pause was unusually long and somewhat awkward, whereas a deliberately planned rhetorical pause might have been expected to be smoother. The new hypothesis then destabilizes itself again and may lead the listener back to `accidental'. In this way, the successive contextualizations associated with the two hypotheses can generate an oscillatory deliberation.

But the very same initial hypothesis can also give rise to a self-reinforcing process. Suppose that, after provisionally selecting `accidental', the listener recalls that the speaker had already shown slight signs of hesitation, had briefly lost eye contact with the audience, or seemed momentarily uncertain about the next words. These contextual elements fit naturally with the hypothesis of an accidental pause and therefore increase its plausibility. Rather than making the alternative answer more attractive, the contextualization produced by the selected answer now provides additional support for that same answer. The working hypothesis is consequently stabilized and may quickly become the final judgment.

It is worth noting that the two regimes also differ with respect to the memory requirements identified in Sec.~\ref{origin-relevance}. A self-reinforcing deliberation terminates after a single contextualization and can be described without keeping track of how the current state was reached, whereas an oscillatory deliberation typically requires precisely such a record, if only to recognize that the same configurations are being revisited. In this sense, the origin degree of freedom of the Aerts-Broekaert-Smets model can be regarded as the minimal ingredient that a formal description of deliberation must possess.

\end{document}